\documentclass[twocolumn, titlepage superscriptaddress]{revtex4-1}
\usepackage{amsmath}
\usepackage{graphicx}
\usepackage{color}

\def\half{{\textstyle \frac{1}{2}}}

\def\Tr#1{{\rm Tr}\left( #1 \right)}
\def\Det#1{{\rm Det}\!\left( #1 \right)}

\begin{document}

\title{Mechanics of invagination and folding: hybridized instabilities\\ when one soft tissue grows on another}
\author{Tuomas Tallinen}
\affiliation{Department of Physics and Nanoscience Center, University of Jyvaskyla, 40014 Jyvaskyla, Finland}
\email{tuomas.tallinen@jyu.fi}
\author{John S. Biggins}
\affiliation{Cavendish Laboratory, Cambridge University, Cambridge, UK}
\email{jsb56@cam.ac.uk}
\date{\today}
\begin{abstract}
We address the folding induced by differential growth in soft layered solids via an elementary model that consists of a soft growing neo-Hookean elastic layer adhered to a deep elastic substrate. As the layer/substrate modulus ratio is varied from above unity towards zero we find a first transition from supercritical smooth folding followed by cusping of the valleys to direct subcritical cusped folding, then another to supercritical cusped folding.  Beyond threshold the high amplitude fold spacing converges to about four layer thicknesses for many modulus ratios. In three dimensions the instability gives rise to a wide variety of morphologies, including almost degenerate zigzag and triple-junction patterns that can coexist when the layer and substrate are of comparable softness. Our study unifies these results providing understanding for the complex and diverse fold morphologies found in biology, including the zigzag precursors to intestinal villi, and disordered zigzags and triple-junctions in mammalian cortex.
 \end{abstract}

\maketitle
   
\section{Introduction}

Some biological tissues have evolved into folded morphologies to accommodate large surfaces in small volumes; folding of the mammalian brain is associated with expansion of its outer cerebral cortex, and the folded epithelium in the gut provides large surface area for absorption of nutrients.  Pattern formation in morphogenesis is typically attributed to chemical instabilities within Turing's reaction-diffusion paradigm \cite{turing1952}, but increasing experimental evidence shows that these folded morphologies actually emerge from  smooth precursors via mechanical instabilities driven by differential growth of layered tissues \cite{xu2010, shyer2013}. Recent  studies have considered mechanical descriptions of brain folding \cite{richman1975, toro2005, bayly2013, kuhl2014, tallinen2014}, epithelial folding in the gut \cite{shyer2013, hannezo2011} and carcinomas \cite{risler2013}, fingerprint formation \cite{kucken2005}, and circumferential buckling in tubular organs and tumors \cite{moulton2011, li2011, dervaux2011}. 

We consider the most elementary model for soft tissue folding: a growing elastic layer with shear modulus $\mu_1$ and relaxed thickness $h$ adhered to an infinite substrate with shear modulus $\mu_2\equiv \mu_1/\eta$. The layer's growth induces compression, ultimately causing it to buckle \cite{dervaux2011, dervaux2011buckling, guvendiren2010solvent, biot1965, cao2012, wang2014,rogers2007, sultan2008, huang2005, audoly2008, cai2011,breid2011, hutchinson2013role, budday2015period, wu2013, allen1969, biot1965}. If $\eta\gg1$ the  layer buckles supercritically into smooth wrinkled patterns \cite{rogers2007, sultan2008, huang2005, audoly2008, cai2011, breid2011} with wavelength $l \sim h \eta^{1/3}\gg h$ explicable via small strains and linear stability analysis \cite{allen1969}. If $\eta \ll 1$, the substrate does not deform, but after sufficient growth the layer's free surface invaginates into cusped furrows called sulci or creases via an exotic supercritical non-linear instability that eludes analytic treatment \cite{tanaka1987, trujillo2008, hohlfeld2011, hohlfeld2012, cai2012, mora2011, tallinen2013}. However, most cases of biological interest involve a growing soft tissue on a comparably soft substrate -- a mechanically intriguing regime \cite{biot1965, dervaux2011buckling, cao2012, hutchinson2013role, budday2015period, wu2013, wang2014, guvendiren2010solvent} that must span a transition from smooth to cusped folding. Previous theoretical studies have focused on two dimensions and have considered post-bifurcation behavior when approaching this transition from the stiff-layer smooth wrinkling side \cite{cao2012, hutchinson2013role, budday2015period}, and the linear stability of circular layers either side of the transition \cite{dervaux2011buckling}. Here we use linear analysis and numerics to explore folding around the transition itself, including beyond threshold behavior in two and three dimensions. We discover the transition occurs via an unexpected region of subcritical cusped folding and is accompanied by an unusually rich set of three-dimensional pattern morphologies, which we argue underpin the mechanics and diversity of invaginations in developmental biology. 

\section{Results}

We examine the stability of a layer occupying $0<z<a$ atop a substrate occupying $z<0$. The layer has undergone growth, swelling, or pre-strain such that, without adhesion, it would undergo a relaxing deformation $G=\mathrm{diag}(g_x,g_y,g_z)$, becoming a flat slab with thickness $h=a/g_z$. We model both tissues as incompressible neo-Hookean elastic solids so that, if this reference state is subject to a displacement $\mathbf{u}$ and corresponding deformation gradient $F=I+\nabla \mathbf{u}$, the elastic energy per unit volume  is $\half \mu_2\Tr{F\cdot F^T}$  in the substrate and $\half \mu_1\Tr{F\cdot G^{-1} \cdot G^{-T}\cdot F^T}$ in the layer, while incompressibility requires $\Det{F}=\Det{G}=1$. 

\begin{figure}[b]
\includegraphics[width=\columnwidth]{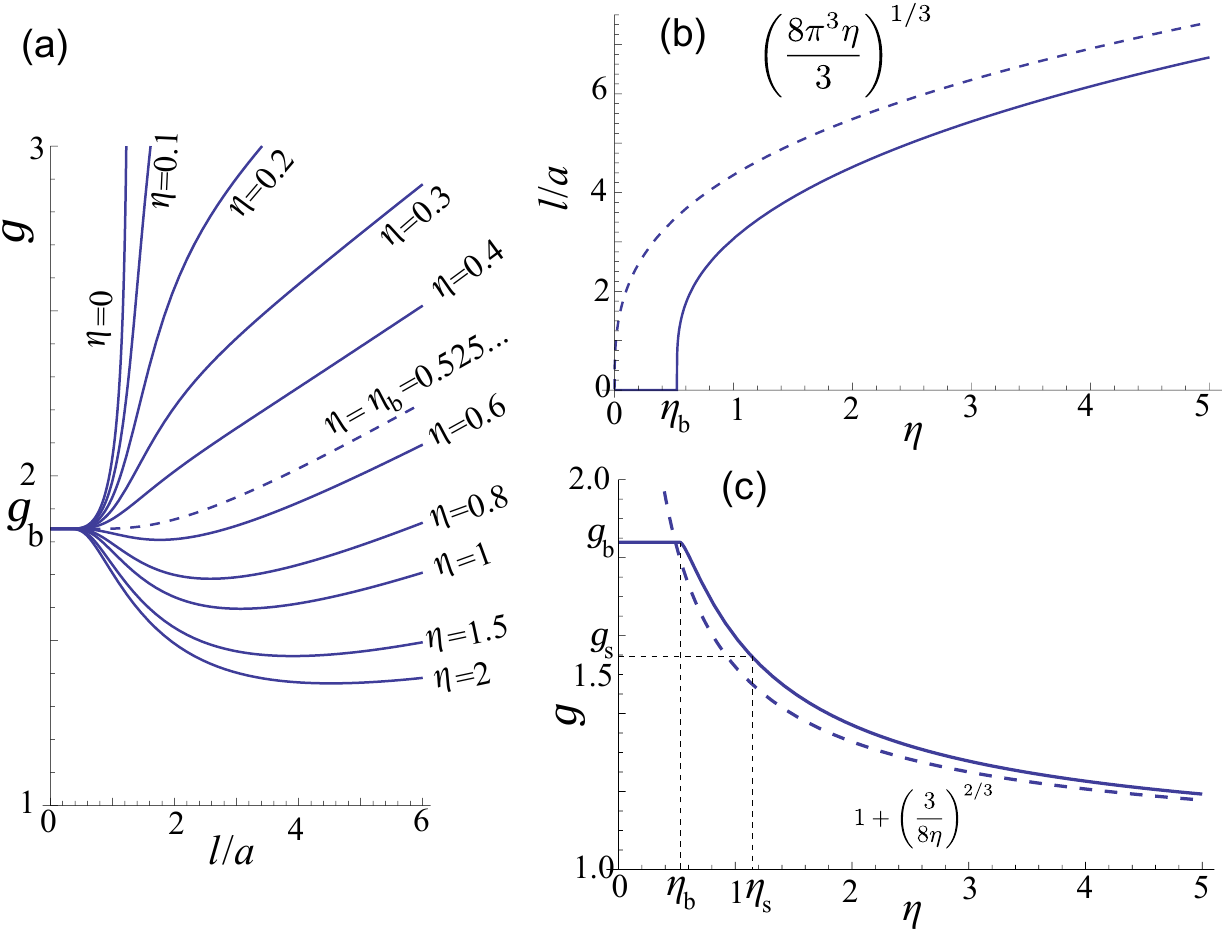}
\caption{ Elastic linear instability of a growing slab on a substrate, showing a transition from finite to zero wavelength instability at modulus ratio $\eta_b\approx0.525$. (a) Growth $g$ required for instability at wavelength $l$ relative to a deformed thickness $a$ for a range of moduli ratios. (b-c) Wavelength and threshold growth of the first unstable mode as a function of $\eta$. Dashed lines indicate the asymptotic results for stiff layers.
}
\label{fig:theory}
\end{figure}

\begin{figure*}[t]
\begin{center}
\includegraphics[width=155mm]{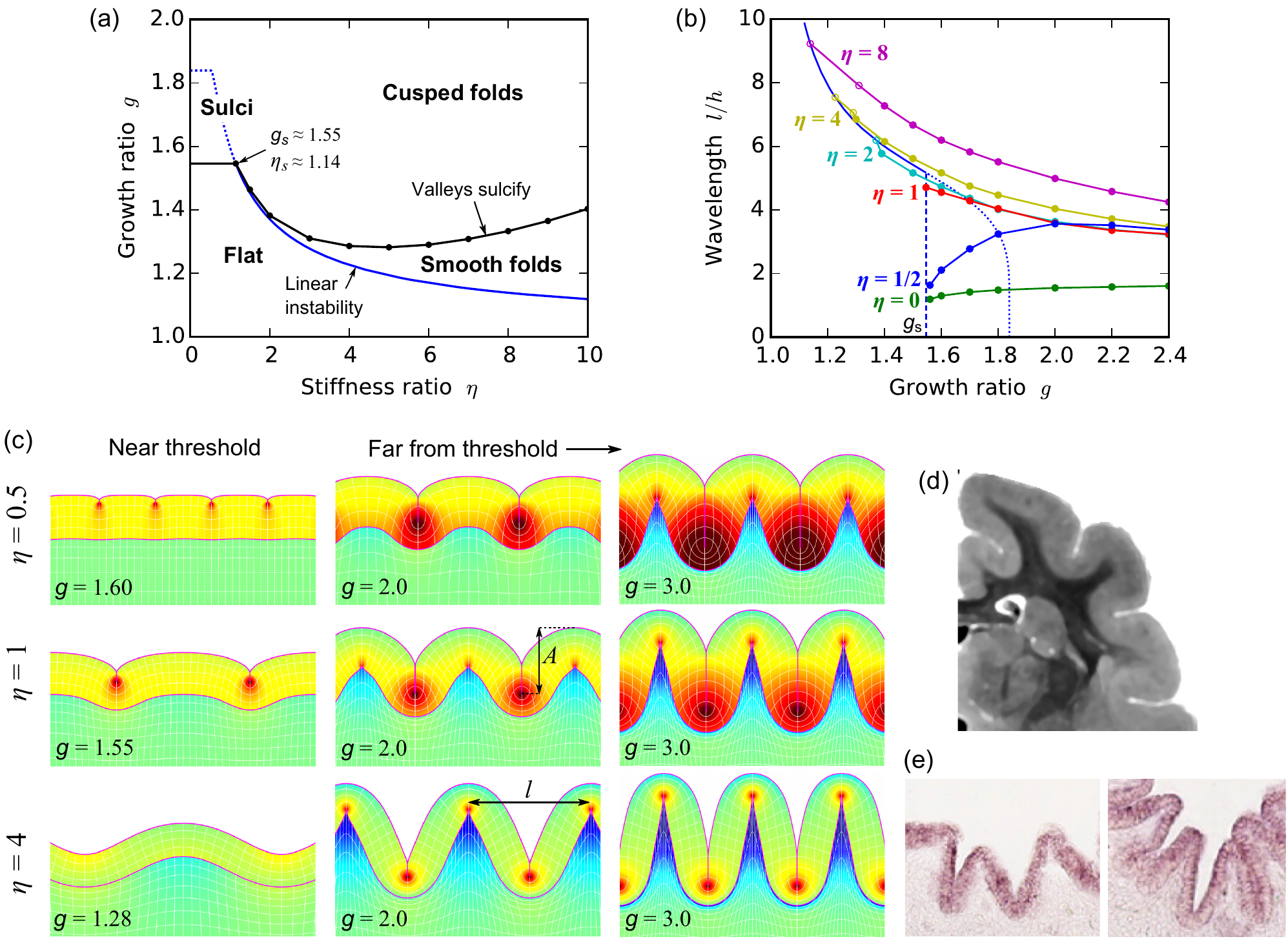}
\caption{ Folding of a uniaxially growing layer in two dimensions.
(a) Morphologies as a function of modulus ratio $\eta$ and growth ratio $g = g_x$. The diagram includes the linear instability curve (blue line, dotted for $\eta < \eta_s \approx 1.14$ where the sulcification threshold is met before the linear instability) and numerically determined thresholds for cusp formation (black line).
(b) Energy minimizing wavelengths as a function of $g$ for a range of moduli ratios $\eta$. Hollow and solid circles correspond to smooth and cusped folds, respectively. The blue curve is the wavelength of linear instability (the dotted end is above the sulcification threshold $g_s \approx 1.55$). 
(c) Simulated minimum energy states near threshold (left column) and far from threshold (middle and right column) for $\eta$ = 0.5, 1 and 4. Color indicates isotropic stress: dark red $\leq -5\mu$, dark blue $\geq 5\mu$ ($\mu = \mu_1$ in the layer and $\mu = \mu_2$ in the substrate). 
(d) Cortical folds in a ferret brain (braincatalogue.org). 
(e) Cross-sections of previllous folds in intestinal epithelium of a chick embryo at days 13 and 14 \cite{shyer2015} (credits: A. E. Shyer).
}
\label{simfig2d}
\end{center}
\end{figure*}

\subsection{Linear instability} 

We first consider uniaxial growth, $G=\mathrm{diag}(g,1,1/g)$, and consider the layer's linear stability to infinitesimal isochoric perturbations, $\mathbf{u}=f(z)\sin(k x) \mathbf{\hat{z}}+(f'(z)/k) \cos(k x) \mathbf{\hat{x}}$. The full calculation is in  Appendix \ref{theoretical_SI}; here we restrict ourselves to an overview. The linearized Euler-Lagrange equations for the solids require $g^4 f^{(4)}-k^2 (g^4+1)f^{(2)}+k^4f=0$ (with $g=1$ in the substrate), solved by  $f(z)=a_4\mathrm{e}^{k z}+a_3 \mathrm{e}^{-k z}+a_2 \mathrm{e}^{ k z/g^2}+a_1 \mathrm{e}^{-k z/g^2}$ in the layer and  $f(z)= \mathrm{e}^{k z}(b_1+z k b_ 2)$ in the substrate. Imposing material continuity at the interface and natural boundary conditions at the free surface and interface we conclude the layer becomes linearly unstable when $g$ satisfies
\begin{equation}
\left|
\left(
\begin{array}{cccc}
 b e^{\frac{- a k}{g^2}}  & b e^{\frac{ a k}{g^2}} & 2 e^{-a k} g^4 & 2 e^{a k} g^4 \\
 -2 g^2 e^{\frac{- a k}{g^2}} & 2 e^{\frac{ a k}{g^2}} g^2 & -b e^{-a k} & b e^{a k} \\
 \frac{b \eta }{2}+1 & \frac{b \eta }{2}-1 & \eta  g^4+g^2 & g^4 \eta -g^2 \\
 \eta +1 & 1-\eta  & \frac{b \eta }{2 g^2}+1 & 1-\frac{b \eta }{2 g^2} \\
\end{array}
\right)
\right|=0,\label{det_eqn}
\end{equation}
where $b=1+g^4$. Fig.\ \ref{fig:theory}a plots this threshold as a function of wavelength for different modulus ratios. Minimizing this threshold over wavelengths gives the wavelength and threshold of the first unstable mode as a function of $\eta$, plotted in Fig.\ \ref{fig:theory}b-c. As expected the instability reproduces the long-wavelength wrinkling limit for $\eta\gg1$ and the zero wavelength Biot instability \cite{biot1965} at $g=g_b=1.839...$ for $\eta\ll1$. The transition from finite to zero wavelength instability happens continuously at 
\begin{equation}
\eta_b=\frac{g_b^4-2 g_b^2+\sqrt{5 g_b^8+8 g_b^6+2 g_b^4+1}+1}{g_b^6+3 g_b^4-g_b^2+1}=0.525...\notag
\end{equation}
However, the zero wavelength Biot instability is never observed: it is always preceded by a point of nonlinear instability to infinitesimal cusped sulci \cite{hohlfeld2011}. In incompressible neo-Hookean materials this occurs at a uniaxial stretch $\lambda_s \approx 0.647$ \cite{hohlfeld2011}, corresponding to $g_s = 1/\lambda_s \approx 1.55$. Thus we expect finite wavelength wrinkling as predicted by linear stability for $\eta>\eta_s \approx 1.14$, where the linear threshold intersects $g_s$ (see Fig.\ \ref{fig:theory}c), and direct sulcification at $g_s$ for $\eta<\eta_s$.

\begin{figure*}
\begin{center}
\includegraphics[width=150mm]{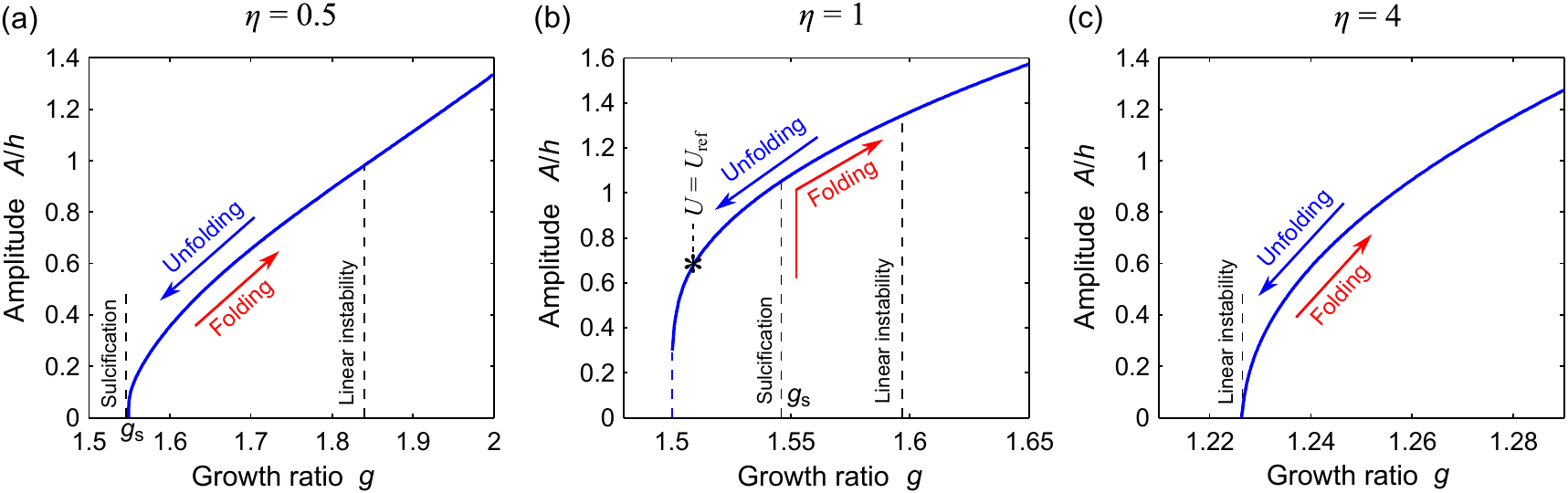}
\caption{Hysteresis loops for three different modulus ratios, demonstrating the three different folding regimes.  (a) When $\eta = 0.5$ sulcification sets in supercritically at $g = g_s \approx 1.55$, producing a non-linear supercritical instability, reminiscent of pure sulcification. (b) When $\eta = 1$ sulcification occurs subcritically at $g=g_s$, immediately producing  a deep cusped fold, while in unfolding the fold also disappears sub-critically at a lower value of $g$. Folding is non-linear and subcritical. (c) When $\eta = 4$ folding is linear and supercritical, reminiscent of pure wrinkling.  
}
\label{unfolding_amplitude}
\end{center}
\end{figure*}

\subsection{Folding in two dimensions}

We verify these predictions and explore folding beyond threshold with numerical simulations (Appendix \ref{numerical_SI}), starting with uniaxial growth in two dimensions. When $\eta > \eta_s$ the surface indeed buckles smoothly at the linear instability threshold. This buckling amplifies surface compression in the valleys, producing the sulcification strain there shortly thereafter.  The valleys then sulcify in a secondary bifurcation which occurs before $g_s$ up to $\eta \approx 10$ (Fig.\ \ref{simfig2d}a).  Cusps form first on the surface and then on the interface between the layer and substrate at high $g$ (Fig.\ \ref{simfig2d}c). However, the interface in real systems may not be strictly sharp, suppressing interfacial cusping.

When $\eta < \eta_s$ sulcification sets in at $g_s$, before the linear instability. For $\eta$ near $\eta_s$ the energy of the system at $g = g_s$ is minimized by a cusped fold that has a finite amplitude (see the layer with $\eta = 1$ and $g = 1.55 \approx g_s$ in Fig.\ \ref{simfig2d}c), implying that sulcification is a nonlinear subcritical instability when $\eta$ is near $\eta_s$. The instability is known to be supercritical in both the stiff layer wrinkling limit $\eta\gg1$ and in the  $\eta \to 0$ pure sulcifications limit \cite{hohlfeld2012, tallinen2013}, so this region of sub-criticality between the two limits is a surprise, stemming from hybridization of the two modes. In Fig.\ \ref{unfolding_amplitude} we show hysteresis loops for the instability at three values of $\eta$ characterizing the three regimes: linear and super-critical for $\eta\gg1$, non-linear and subcritical for $\eta \approx 1$ and non-linear and supercritical for $\eta\lesssim 0.5$. Analogous subcritical instabilities were recently identified in two-dimensional elastic systems with jointly compressed layer and substrate \cite{diab2013, jin2015}. These systems differ from ours particularly strongly in the region of subcriticality: indeed at $\eta=1$ the distinction between layer and substrate vanishes in the jointly compressed system, so the wavelength scales with the total depth of the system, whereas in our system the layer is distinguished by its growth even at $\eta=1$ and the wavelength is always finite despite the substrate being infinite.

We optimized the wavelengths of the folds for minimum energy (Fig.\ \ref{simfig2d}b) finding that in the smooth-folding regime, $\eta > \eta_s$, the wavelength $\l$ decreases past threshold (in Eulerian coordinates), which is similar to folding of stiff layers \cite{rogers2007}. In contrast, when $\eta < \eta_s$ the optimal wavelength increases past threshold. Consequently the optimal wavelength settles to about four times the undeformed layer thickness over a wide range of $\eta$ when $g$ becomes large. This agrees with the folding patterns in the brain and the gut \cite{shyer2013, tallinen2014}, and is a hallmark of mechanical compressive folding. 

The low $\eta$ folds are notably asymmetric, with a deeply cusped upper surface and a wavy bottom surface, in contrast to nearly top-bottom symmetric folds at $\eta > \eta_s$ that take a sawtooth-like form at high amplitudes (Fig.\ \ref{simfig2d}c). This agrees again with the observations in folded tissues; the brain has a low modulus ratio between the cortex and sublayers ($\eta \approx 1$ \cite{xu2010}) leading to thick and deeply cusped asymmetric folds (Fig.\ \ref{simfig2d}d) \cite{bok1959}, whereas narrower folds with a sawtooth profile (Fig.\ \ref{simfig2d}e) are seen during gut development, owing to a relatively stiff epithelium that folds on a softer mesenchyme \cite{shyer2013, shyer2015}. We note, however, that in a developing chick gut the epithelium and mesenchyme are both confined by outer layers of smooth muscle, which contributes to folding at early stages \cite{shyer2013}.

\subsection{Folding in three dimensions}
 
If $G=\mathrm{diag}(g_x,g_y,1/(g_x g_y))$ is not uniaxial, but we nevertheless consider $x-z$ plane-strain deformations then, as seen in eqn.\ \ref{plane-strain-energy-density},  the elastic energy depends on two dimensionless parameters, $\eta/g_x^2$ and $\eta g_x^2 g_y^2$. Our plane-strain calculations thus hold with the identifications $\eta\to g_y\eta$ and $g\to g_x \sqrt{g_y}$. In particular, for transversely isotropic growth, $g=g_x=g_y$, the sulcification threshold is at $g_x \sqrt{g_y} = g^{3/2} \approx 1.55$, i.e, $g_s \approx 1.34$, and the linear threshold intersects $g_s$ at $\eta_s \approx 0.86$. This plane-strain calculation captures the full linear stability of the system since the Fourier modes that would make up a truly three-dimensional pattern cannot interact at the level of linear stability. 

Before examining folded states in three dimensions, we recall previous results in the stiff and soft layer limits: In stiff layers ($\eta \gg \eta_s$) a hexagonally ordered pattern of dents forms near threshold \cite{cai2011, breid2011} (although analyses based on the plate theory predict a checkerboard pattern in the stiff film limit \cite{audoly2008, cai2011}) and a zigzag (herringbone) pattern is the preferred far-from-threshold pattern \cite{cai2011, breid2011}. In contrast, in clamped soft layers ($\eta = 0$) we previously identified a square pattern of line-like sulci with alternating orientations as the optimal near threshold pattern and a hexagonal array of triple-junction sulci as the optimal far-from-threshold pattern \cite{tallinen2013}. There should thus be transitions between the two sets of patterns as the layer stiffness is reduced from large toward zero. 

Whereas folding of relatively stiff layers can be studied even at large overstresses using experimental setups based on swelling elastomers \cite{cai2011, breid2011}, folding of soft layers is harder to realize in experimental model systems because large swelling ratios are needed to reach far-from-threshold states. Patterns of line-like sulci and triple-junction sulci have been produced experimentally in soft layers adhered to rigid substrates \cite{tanaka1987, trujillo2008, guvendiren2010solvent}, but very few experiments have explored three dimensional folding patterns in the regime of modest stiffness ratios \cite{guvendiren2010solvent, tallinen2014}.

\begin{figure*}[t!]
\begin{center}
\includegraphics[width=165mm]{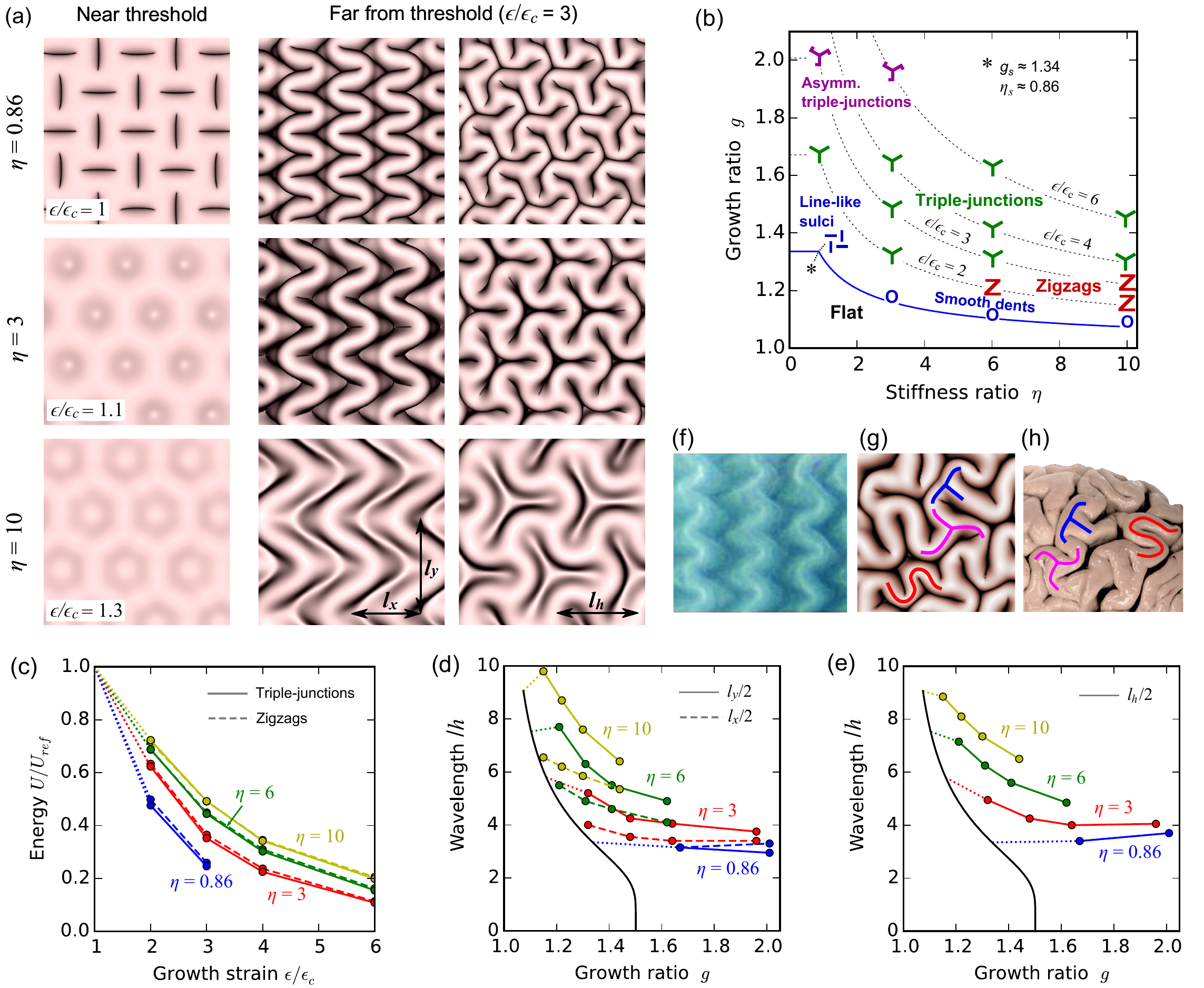}
\caption{ Three dimensional folding of an equibiaxially growing layer.
(a) Energy minimizing periodic patterns near threshold (left) and far from the threshold (right, zigzag and triple-junction patterns are shown for each $\eta$).
(b) Diagram of energy minimizing ordered patterns as a function of $\eta$ and $g$. The symbols (Z's for zigzags, Y's for triple-junctions, o's for smooth dents, etc.) indicate the minimum energy pattern at each simulated point. The blue curve is the threshold for linear instability or sulcification. The dashed lines are contours for growth strains $\epsilon/\epsilon_c$.
(c) Deformation energies of zigzags and triple-junctions relative to the energies of flat reference states.
(d,e) Energetically optimal half-wavelengths [$l_x$, $l_y$, and $l_h$ are indicated in (a)] of zigzags and triple-junctions, respectively. Thick black curves indicate the wavelength of linear-instability.
(f) Zigzags in epithelium of a developing chick gut \cite{shyer2013}. 
(g) A simulated soft layer growing in a large simulation domain ($\eta = 1$ and $g = g_x = g_ y = 2$) folds spontaneously into a disordered mixture of Y and T-shaped triple-junction folds and S-shaped zigzag folds, reminiscent of the folded cerebral cortex (h) \cite{bradbury2005}. 
}
\label{simfig3d}
\end{center}
\end{figure*}

We simulate three dimensional folding patterns induced by transversely isotropic growth of the layer on periodic domains that contain one unit cell of the pattern with energetically optimized wavelengths (Appendix A). We use external guiding forces to inititate the patterns that, once formed, are stable in the neighborhood of $\eta$,$g$-space where they are the energy minimizing patterns. Optimal near-threshold patterns, including a rectangular array of line-like sulci at $\eta \lesssim \eta_s$ and smooth hexagonal dents at $\eta > \eta_s$, are shown on the left of Fig.\ \ref{simfig3d}a. These patterns minimize the energy only at $\epsilon/\epsilon_c$ close to unity, where we define the relative growth strain $\epsilon/\epsilon_c = \frac{g - 1}{g_c-1}$ to conveniently quantify distance from the threshold ($\epsilon = g - 1$ and $\epsilon_c = g_c - 1$ where $g_c$ is the growth ratio at the threshold). As in two dimensions, folding is subcritical when $\eta$ is near $\eta_s$, evidenced by a finite amplitude pattern at the threshold $g = g_s$ when $\eta = \eta_s$ (Fig.\ \ref{simfig3d}a). For far-from-threshold states ($\epsilon/\epsilon_c \geq 2$) we construct zigzag patterns as well as patterns of hexagonal triple-junctions, discovering that these two patterns are almost equally good energy minimizers (Fig.\ \ref{simfig3d}c). However, the triple-junctions are slightly better at low $\eta$ and/or high $\epsilon/\epsilon_c$, while zigzags become relatively better toward large $\eta$, as expected. Fig.\ \ref{simfig3d}b shows a phase diagram summarizing these findings. 

The optimal wavelengths of the zigzags and triple-junctions are shown in Fig.\ \ref{simfig3d}d and \ref{simfig3d}e, respectively. As in the uniaxial case, the wavelengths decrease with $g$ when $\eta > \eta_s$, but remain nearly steady when $\eta \approx \eta_s$, resulting in an approximate collapse of the wavelengths toward large $g$ such that the width of the folds is again roughly four times the undeformed layer thickness. This means that further expansion of the layer is accomodated by increasing amplitude. The optimal $x,y$-aspect ratio of zigzags decreases with decreasing $\eta$ such that the zigzags in soft layers are relatively shorter than their stiff-layer counterparts \cite{cai2011}.

Given the nearly equal energies of the zigzag and triple-junction folds, it is not surprising that systems with modest modulus ratios, such as the folded cerebral cortex (Fig.\ \ref{simfig3d}h) or its gel layer mimics \cite{tallinen2014}, show disordered mixtures of zigzags and triple junctions (in contrast, epithelium of the gut shows ordered zigzags, see Fig.\ \ref{simfig3d}f, owing to a higher modulus ratio and anisotropy \cite{shyer2013}). By simulating unaided folding of the biaxially growing layer in a large simulation domain whose width is multiple times the wavelength of the patterns (Appendix A), we see comparatively ordered near-threshold patterns similar to those in Fig.\ \ref{simfig3d}a, but the far-from-threshold patterns appear as disordered mixtures of zigzag-folds and triple-junctions, reminiscent of the folded cerebral cortex when $\eta \approx 1$ (Figs.\ \ref{simfig3d}f and \ref{simfig3d}g). The disordered patterns include some T-shaped triple-junction folds that lack the rotational symmetry of the Y-shaped ones. T-shaped folds do not, however, form a simple periodic pattern and they may thus be considered as variants of the ideal Y-shaped folds. We note that the disordered patterns resulting from unaided folding have slightly higher energies than the corresponding periodic zigzag or triple-junction patterns. 

\section{Summary}

Our study demonstrates that complex fold patterns emerge from even the simplest systems incorporating elasticity and differential growth, providing a guide for more specific applications incorporating, for example, curved geometries, anisotropy, prestressed substrates, compressibility and stress relaxation. We show a transition as a function of modulus ratio from supercritical smooth finite-wavelength buckling followed by sulcification of valleys to direct strongly subcritical sulcification at $0.86<\eta_s<1.14$, depending on growth anistropy. This is accompanied by a transition from  contraction to  elongation of past threshold wavelengths, driving a convergence to spacings within three to four layer thicknesses over a range of modulus ratios in both two and three dimensions. In equibiaxial growth there is also a beyond threshold transition from zigzags to triple-junction folds within the typical range of modulus ratios in soft layered tissues, underpinning the morphogenetic diversity of mechanical folding.

\appendix

\section{Details of numerical simulations}\label{numerical_SI}
The simulations are based on the explicit finite element method with discretizations to constant strain triangle (2D) or tetrahedron (3D) elements. The elements have a quasi-incompressible nodal pressure formulation with a bulk modulus $K = 10^3 \mu$ (2D) or $K = 10^2 \mu$ (3D). The simulated substrate has a thickness ten times the layer thickness. Whether the substrate has a clamped or free base has a negligible effect on the results, confirming that it provides a good approximation of an infinite substrate. 

In 2D plane-strain simulations the domain includes a half fold with symmetric boundary conditions at sides. The domain is discretized to a rectangular mesh of triangles such that the growing layer includes 200 elements through its thickness. A fold is initiated by applying a small normal force to the surface at the edge of the domain. The surface is prevented to cross the vertical line that defines the boundary of the domain to prevent self-intersection of the fold. The energetically optimal wavelength is searched by adjusting the relative width of the domain.

In 3D simulations the domain has periodic boundary conditions at the lateral boundaries. The domain is discretized to tetrahedrons such that the nodes form a hexagonal prism mesh. The growing layer includes 10 elements through its thickness. Self-contacts of the surface are accounted for by preventing nodes penetrating faces at the surface. In the simulations of periodic patterns with a unit cell-wide domains point forces are applied in selected locations of the domain to initiate formation of the patterns. When the dimensions of the domain are close to the optimal wavelength of the desired pattern, a stable periodic pattern forms easily. Dimensions of the domain are fine-tuned to find the energy minimizing wavelength. In the simulations of large rectangular domains (Fig.\ \ref{simfig3d}f) the domain has a width 30 times the layer thickness and the layer is allowed to fold and relax freely. The surface of the layer in this case has small random perturbations that allow folding to initiate.

The simulation widths for the hysteresis loops in Fig.\ \ref{unfolding_amplitude} were $2.50h$ for $\eta=0.5$, $4.71 h$ for $\eta=1$ (corresponding to the energy minimizing wavelength at $g=g_s$) and $7.54 h$ for $\eta=4$ (corresponding to the wavelength of linear instability). The hysteresis loops were calculated in unloading. The loading curves are deduced from the fact we know the system simply becomes unstable at the lower of the linear stability threshold or $g_s$.

\section{Details of linear stability analysis}\label{theoretical_SI}
The total elastic energy of the layer and substrate is
\begin{equation}
E=\int_{-\infty}^{\infty}\int_{-\infty}^{\infty}\int_{-\infty}^a\half \mu \Tr{F\cdot G^{-1}\cdot G^{-T} \cdot{F}^T}\mathrm{d}z \mathrm{d}x\mathrm{d}y,
\end{equation}
with
\begin{equation}
(\mu,G)=\left\{
     \begin{array}{lr}
       (\eta,  \mathrm{diag}(g_x, g_y, 1/(g_x g_y)))&\mathrm{\ \ \ }0<z<a\\
       (1,I)  &\mathrm{\ \ \ } z<0
     \end{array}
   \right.      \end{equation}
If we restrict attention to $x-z$ plane-strain deformations then the above integrand can be written out in terms of components of $F$ as 
     \begin{align}
\left\{
     \begin{array}{lr}
      \!\! c_1\!+\!\half \eta \left(g_x^2 g_y^2 \left(F_{xz}^2+F_{zz}^2\right)\!+\!g_x^{-2}(F_{xx}^2+F_{zx}^2)\right)\!&0<z<a\\
      \!\!c_2\!+  \half  \left( F_{xz}^2+F_{zz}+F_{xx}^2+F_{zx}^2\right)  & z<0
     \end{array}
   \right.\label{plane-strain-energy-density}   \end{align}
where the constants $c_1$ and $c_2$  do not depend on $F$, and hence do not enter into energy  minimization.

We enforce incompressibility, encoded by $\Det{F}=1$, by introducing a Lagrange multiplier pressure field $P$, leading to the effective energy density\begin{equation}
L=\half \mu \Tr{F\cdot G^{-1}\cdot G^{-T} \cdot{F}^T}-\mu P (\Det{F}-1).\end{equation}
Minimizing this energy with respect to variations in $\mathbf{u}$ and $P$ gives the Euler-Lagrange equations
\begin{align}
\nabla \cdot \sigma &=0 \label{eqn:stress_div}\\
\Det{F}&=1\label{eqn:det_cont},
\end{align}
where $\sigma=\partial L/\partial \nabla \mathbf{u}$ is the stress-tensor, given by
\begin{equation}
\sigma=\mu \left( F\cdot G^{-1}\cdot G^{-T}-P F^{-T}\right).
\end{equation}

Energy minimization also requires the free surface to be stress free, and the layer and substrate stresses to match at their interface:
\begin{equation}
\sigma\cdot\mathbf{\hat{z}}|_{z=a}=0,\mathrm{\ \ \ \ \ }\sigma\cdot\mathbf{\hat{z}}|_{z=0^{+}}=\sigma\cdot\mathbf{\hat{z}}|_{z=0^{-}}.
\end{equation}

We  restrict our attention to growth in the $x$ direction, $g_x=g$, $g_y=1$, for which we expect buckling in the $x-z$ plane so we consider the linear stability of the flat  state to infinitesimal sinusoidal perturbations with wavenumber $k$, writing the fields as
\begin{align}
P&=P_0+\epsilon \sin(k x) P_1 (z) \\ \mathbf{u}&=
       \epsilon( \cos(k x) d(z)\mathbf{\hat{x}}+ \sin(k x) f(z)\mathbf{\hat{z}}).\notag\end{align}
Working in an $x-z$ basis, to first order in $\epsilon$ we thus have
\begin{align}
F&=I+\epsilon\left(
\begin{array}{cc}
 -k d(z) \sin (k x) & \cos (k x) d'(z) \\
 k \cos (k x) f(z) & \sin (k x) f'(z) \\
\end{array}
\right),\\F^{-T}&=I-\epsilon\left(
\begin{array}{cc}
 -k d(z) \sin (k x) & k \cos (k x) f(z) \\
 \cos (k x) d'(z) & \sin (k x) f'(z) \\
\end{array}
\right).\notag
\end{align}
 At zeroth order in $\epsilon$ we have $F=F^{-T}=I$ and hence piecewise constant stress $\sigma=\mu \left( G^{-1}\cdot G^{-T}-P_0 I\right)$.  The bulk equations are thus already satisfied, and the boundary conditions require $\sigma\cdot \mathbf{\hat{z}}=0$ throughout, solved by
 \begin{equation}
 P_0= \mathbf{\hat{z}\cdot}G^{-1}\cdot G^{-T}\cdot{ \mathbf{\hat{z}}}=\left\{
     \begin{array}{lr}
       g^2 &\mathrm{\ \ \ }0<z<a\\
       1  &\mathrm{\ \ \ } z<0.
     \end{array}
   \right.
 \end{equation}
Expanding eqn.\ \ref{eqn:det_cont} to first order in $\epsilon$ reduces it to $\nabla \cdot \mathbf{u}=0$, requiring  $d(z)=f(z)/k$. In the layer, eqn.\ \ref{eqn:stress_div} expanded to first order requires
\begin{align}
k^2 \left(f'(z)+g^2 P_1(z)\right)-g^4 f'''(z)&=0\\
-g^4 f''(z)+k^2 f(z)+g^2 P_1'(z)&=0.
\end{align}
The first of these can be solved algebraically for $P_1$, giving
\begin{equation}
P_1(z)= \frac{g^2 f^{(3)}(z)}{k^2}-\frac{f'(z)}{g^2},
\end{equation}
and substituting this into the second gives the governing equation for $f(z)$:
\begin{equation}
g^4 f^{(4)}(z)-\left(g^4+1\right) k^2 f''(z)+k^4 f(z)=0.
\end{equation}
The same results hold in the substrate, but with $g=1$. The solution for $f$ in the layer and the substrate is thus
\begin{equation}
f(z)\!=\!\left\{
     \begin{array}{lr}
     \!\! a_4\mathrm{e}^{k z}\!+\!a_3 \mathrm{e}^{-k z}\!+\!a_2 \mathrm{e}^{ k z/g^2}\!+\!a_1 \mathrm{e}^{-k z/g^2} &\mathrm{\ }0<z<a\\
      \!\! \mathrm{e}^{k z}(b_1+z k b_ 2)  &\mathrm{\ \ \ } z<0,
     \end{array}\right.
\end{equation}
where the $a_i$ and $b_i$ are constants of integration, and we have disgarded the unbounded  solutions in the substrate.  Continuity of $\mathbf{z}$ displacement at the interface requires continuity of $f_2$ at $z=0$, giving
\begin{equation}
b_1=a_1+a_2+a_3+a_4\end{equation}
while continuity of $\mathbf{x}$ displacement requires continuity of $f_2'$ at $z=0$, giving
\begin{equation}
b_2=(a_2-a_1)/g^2-(2a_3+a_2+a_1).\end{equation}
The first order correction to the free boundary conditions at $z=a$ requires
\begin{align}
&a_1 \left(g^4+1\right)+a_2 \left(g^4+1\right) e^{\frac{2 a k}{g^2}}+2 a_3 g^4 e^{a \left(\frac{1}{g^2}-1\right) k}\notag \\&+2 a_4 g^4 e^{a \left(\frac{1}{g^2}+1\right) k}=0\\
&-2 a_1 g^2+2 a_2 g^2 e^{\frac{2 a k}{g^2}}-a_3 \left(g^4+1\right) e^{a \left(\frac{1}{g^2}-1\right) k}\notag \\&+a_4 \left(g^4+1\right) e^{a \left(\frac{1}{g^2}+1\right) k}=0,
\end{align}
and the  first order correction to the interfacial boundary conditions at $z=0$ require
\begin{align}
&a_1 \left(\eta  g^4+\eta +2\right)+a_2 \left(\eta  g^4+\eta -2\right)+2 a_3 \left(\eta  g^4+g^2\right)\notag \\&+2 a_4 g^2 \left(g^2 \eta -1\right)=0\\
&2 a_1 g^2 (\eta +1)-2 a_2 g^2 (\eta -1)+a_3 \left(\eta  g^4+2 g^2+\eta \right)\notag \\&+a_4 \left(2 g^2-\left(g^4+1\right) \eta \right)=0.
\end{align}
These four linear equations can be re-expressed  as single matrix equation of the form $M\cdot \left(a_1,a_2,a_3,a_4 \right)=0$, which will only admit non-trivial solutions if $\Det{M}=0$,  the condition displayed in eqn.\ \ref{det_eqn}. We find the first point of instability by finding the wave-number that satisfies this condition with the minimum $g$. Solutions to this eqn are shown for a variety of different moduli ratio in fig.\ 1a in the main text. Each line shows the strain required for instability at each wavelength for a given slab. The actual point and wavelength of instability are given by the first unstable mode, that is, by the minimum of the curve. Inspecting fig.\ 1a we see that all layers are unstable to a zero wavelength instability at $g=g_b=1.839...$, which is the Biot strain. We can extract this from the determinant condition by noting that it contains two terms which diverge at small wavelength:
\begin{equation}
\left(g^6-3 g^4-g^2-1\right) A(g,\eta) e^{ak+\frac{a k}{g^2}}+B(g,\eta) e^{ak -\frac{ak}{g^2}}=0,
\end{equation}
where 
\begin{align}
&A=\left(g^2-1\right)^2\\ &\times \left(g^6  \eta^2-g^4  \eta (3  \eta+2)-g^2 ( \eta+2)^2- \eta ( \eta+2)\right)\notag \\ 
&B=-\left(g^2+1\right)^2  \left(g^6+3 g^4-g^2+1\right) \\ &\times \left(g^6 \eta^2+g^4 \eta (3 \eta-2)-g^2 (\eta-2)^2+(\eta-2) \eta\right).\notag
\end{align}
 For asymptotically small wavelength the left exponential dominates and, since $A\ne0$ the condition is only satisfied at the Biot strain $g_b=\frac{1}{3} \left(\sqrt[3]{19+3 \sqrt{33}}+\sqrt[3]{19-3 \sqrt{33}}+1\right)=1.839...$  where $g^6-3 g^4-g^2-1=0$. Further inspecting fig.\ 1a we see that, for sufficiently soft layers ($\eta\le0.525...$) this zero wavelength Biot solution is the first unstable mode, whereas for stiffer layers there is a finite wavelength buckling at a lower growth, giving rise to a more conventional buckling. We characterize the instability by minimizing the curves in fig.1a from the main text over wavelength to find the first unstable mode for each modulus ratio, which gives the wavelength of instability and the growth required. These are plotted against modulus ratio in figs.\ 1b-c in the main paper, which again clearly show the transition from Biot zero-wavelength behavior to finite wavelength buckling at  $\eta\ge0.525...$.
  
For layers marginally stiffer than $\eta_b$ the threshold for instability is just below $g_b$ so we can analyze the instability by setting $g=g_b+\epsilon$ and considering the leading behavior in $\epsilon$. The Biot polynomial then becomes $g^6-3 g^4-g^2-1=\left(6 g_b^5-12 g_b^3-2 g_b\right) \epsilon$ so the condition for instability becomes
\begin{equation}
\left(6 g_b^5-12 g_b^3-2 g_b\right) \epsilon A(g_b,\eta) e^{ak+\frac{a k}{g_b^2}}+B(g_b,\eta) e^{ak -\frac{ak}{g_b^2}}=0,
\end{equation}
which is solved by
\begin{equation}
\epsilon=\frac{B(g_b,\eta) e^{ -\frac{2ak}{g_b^2}}}{\left(6 g_b^5-12 g_b^3-2 g_b\right)  A(g_b,\eta) }.
\end{equation}
This small correction to the Biot-threshold for short but non-zero wavelengths is positive, indicating the Biot point is a minima in fig.\ 1a, if $B(g_b,\eta)>0$ and negative, indicating the Biot point is a maxima, if $B(g_b,\eta)<0$. The crossover occurs at  $B(g_b,\eta_b)=0$ which, inspecting the form of $B$, requires $g_b^6 (\eta_b)^2+g_b^4 \eta_b (3 \eta_b-2)-g_b^2 (\eta_b-2)^2+(\eta_b-2) \eta_b=0$, satisfied if
\begin{equation}
\eta_b=\frac{g_b^4-2 g_b^2+\sqrt{5 g_b^8+8 g_b^6+2 g_b^4+1}+1}{g_b^6+3 g_b^4-g_b^2+1}=0.525...
\end{equation}
Further analysis indicates that the transition is continuous but logarithmic, in the sense that the unstable wavelength vanishes as $\lambda \sim 1/\log(\eta-\eta_b)$. \\

\acknowledgements
T. T. acknowledges the Academy of Finland for funding. The computational resources were provided by CSC -- IT Center for Science.

\end{document}